\documentclass{ceurart}
\usepackage{silence}
\usepackage{amsmath,amssymb}
\usepackage{booktabs}
\usepackage{multirow}
\usepackage{multicol}
\usepackage{enumitem}

\title{Submitted and Diagnostic Analysis of Full-Text Temporal Retrieval for LongEval-Sci}
\title[mode=sub]{LongEval-Sci at CLEF 2026}

\begin{document}

\copyrightyear{2026}
\copyrightclause{Copyright for this paper by its authors.
  Use permitted under Creative Commons License Attribution 4.0
  International (CC BY 4.0).}

\conference{CLEF 2026 Working Notes, 21 -- 24 September 2026, Jena, Germany}

\author[1]{Haijian Wu}[%
orcid=0009-0007-0311-6235,
email=willwuhj@gmail.com,
url=https://wijowill.github.io/,
]
\cormark[1]
\fnmark[1]
\author[1]{Yingdong Yang}[%
orcid=0009-0007-5546-5293,
email=yingdongyang0305@outlook.com,
url=https://github.com/yyd859/,
]
\fnmark[1]
\cortext[1]{Corresponding author.}
\fntext[1]{These authors contributed equally and share first authorship.}

\begin{abstract}
LongEval-Sci evaluates scientific retrieval under collection change, where a system should be effective on the current corpus and remain usable as documents accumulate over time. This paper reports both official Task 1 results and development diagnostics for LongEval-Sci 2026. We compare the official PyTerrier BM25 and Qwen3 dense baselines with full-text BM25, additive and router variants, temporal full-text retrieval, temporal+citation retrieval, RM3 query expansion, cross-encoder reranking, and reciprocal rank fusion (RRF). In the official DCTR evaluation, the temporalized full-text runs are our strongest submissions: FT BM25+temporal and FT BM25+temporal+citation obtain the best ARP on all three snapshots (0.285, 0.267, and 0.180 nDCG@10) and reduce snapshot-3 relative change from 0.481 for the BM25 pivot to 0.368. Citation features match the temporal-only variant but do not provide a measurable additional gain in the official summary. Our internal snapshot-1 diagnostics show a complementary pattern: full-text BM25 is the strongest single development retriever (DCTR nDCG@10 = 0.3302, MAP = 0.2853), RRF gives the best deep recall (Recall@1000 = 0.9667), and some uncalibrated overlays can sharply degrade top-rank quality. We therefore conclude that full-text retrieval is the strongest foundation, temporal integration can improve official longitudinal effectiveness when applied to that foundation, and citation evidence still requires cleaner ablation and calibration. Beyond ranking, we also report a qualitative weekly IR-system update-monitoring analysis based on ingestion velocity and stale-coverage drift.
\end{abstract}


\maketitle

\section{Introduction}

Scientific search collections are not static. New papers are published, metadata changes, citation links accumulate, and terminology shifts as fields mature. A retrieval system that is tuned once may remain strong under one snapshot yet become less reliable as the collection changes. This is the core motivation of LongEval, which frames information retrieval as a longitudinal evaluation problem rather than a one-time ranking task \cite{cancellieri_longeval_2025_clean,lncs_overview_longeval_2026}.

We study LongEval-Sci 2026 Task 1 from both a ranking and a maintenance perspective. Our main validated ranking result is that temporalized full-text retrieval performs best among our official submissions, while the snapshot-1 development experiments explain why this result should be interpreted as calibrated use of temporal evidence rather than as a general license to boost all recent or temporally favored documents. In particular, we ask: (i) which submitted systems are strongest in the official three-snapshot DCTR evaluation; (ii) which reusable lexical, dense, expansion, reranking, and fusion baselines are strongest in snapshot-1 development diagnostics; (iii) whether publication-time and citation overlays add value beyond full-text lexical retrieval; and (iv) what simple collection-drift signals could indicate that an IR system should be refreshed.

Our contributions are threefold. First, we analyze the official Task 1 results and show that FT BM25+temporal and FT BM25+temporal+citation achieve the strongest ARP and best robustness among our submitted models. Second, we compare official title+abstract BM25 and Qwen3 dense baselines with full-text BM25, RM3, cross-encoder reranking, and RRF fusion in a snapshot-1 development protocol with whole-train and cumulative monthly diagnostics. Third, we report a qualitative weekly update-monitoring policy based on ingestion velocity and stale-coverage drift, while explicitly treating it as a maintenance diagnostic rather than a ranking improvement claim.

\section{Related Work}

\subsection{Longitudinal IR and LongEval}

LongEval evaluates how retrieval systems behave when test data diverges temporally from training data \cite{cancellieri_longeval_2025_clean,ceur_overview_longeval_2026}. This framing is especially relevant for scientific search, where document availability, terminology, and evidence signals such as citations all evolve. We follow this direction by separating absolute effectiveness from robustness under simulated collection growth.

\subsection{Scientific Retrieval Baselines}

The LongEval-Sci baseline space includes sparse retrieval, dense retrieval, and reranking. BM25 remains a strong and efficient foundation for large text collections \cite{zobel_inverted_2006}. Dense retrieval and cross-encoder reranking have also been explored in LongEval-Sci participant systems \cite{agh2025}. In our implementation, this motivates a two-view design: title+abstract for compact lexical and dense retrieval, and full text for a stronger sparse retrieval baseline.

\subsection{Fusion, Query Expansion, and Reranking}

RRF combines ranked lists by rank position rather than raw score, making it useful for merging sparse and dense runs whose scores are not directly comparable \cite{cormack2009rrf}. RM3-style pseudo-relevance feedback expands short queries with terms estimated from initially retrieved documents, addressing vocabulary mismatch \cite{lavrenko2001relevance}. OpenWebSearch LongEval systems similarly emphasize incremental improvements over existing rankings through reformulation, clustering, reranking, and historical evidence \cite{alexander2024openwebsearch,alexander2025openwebsearch}. Our fusion layer follows this philosophy: it operates over existing run files and does not create a new index.

\subsection{Temporal and Citation Evidence}

Temporal retrieval work has shown that freshness should be balanced with relevance in a query-dependent way \cite{dai_learning_2011}. TempRALM makes a related argument for temporally aware retrieval-augmented language models, adding temporal relevance without replacing the underlying retrieval index \cite{gade_tempralm_2025}. Citation-based scientific search provides another signal family: citation links, co-citation, and citation-based ranking can help identify relevant scientific literature beyond lexical matching alone \cite{janssens2015cocites,janssens2020validation,belter2017ranking,bascur2023citationclusters}. Our experiments treat these signals as reranking-time overlays, not as new first-stage indexes.

\subsection{Dynamic Index Maintenance}

The long-term update question is connected to dynamic indexing. Incremental inverted indexing and immediate-access dynamic indexing study how text indexes can support growing collections without full rebuilds \cite{brown_fast_nodate,asadi_fast_2013,moffat_efficient_2023}. Streaming vector-index maintenance studies similar problems for dense search \cite{mohoney_incremental_2024}. We treat update triggers as a separate system-maintenance layer rather than mixing them into the ranking-model discussion.

\section{Task, Data, and Protocol}

We use LongEval-Sci 2026 Task 1 through the local \texttt{ir-datasets-longeval} cache and the official Task 1 evaluation report. The official results use discretized Document Click-Through Rate (DCTR) relevance labels and report nDCG@10-based longitudinal measures across three snapshots. For development diagnostics, we use the \texttt{snapshot-1} train collection. It contains 869,902 documents and 100 training queries. We evaluate primarily with DCTR qrels (8,772 qrel rows) and also report raw qrels (1,183 qrel rows) for sensitivity analysis. Unless otherwise stated, tables in the main text use DCTR qrels.

The official snapshot ranges are March-May 2025 for snapshot-1, June-August 2025 for snapshot-2, and September-November 2025 for snapshot-3. Because training qrels are available for snapshot-1, we use snapshot-1 for model development and construct internal month-filtered views for robustness analysis.

\subsection{Official Evaluation and Model Names}

The official evaluation summarizes each model using Average Retrieval Performance (ARP), Relative Change (RC), and Delta Relative Improvement (DRI). ARP is the retrieval effectiveness at a given snapshot; in the official summary used here, ARP is nDCG@10. RC measures change from snapshot 1 to a later snapshot, normalized by snapshot-1 effectiveness:
\[
\mathrm{RC}_{1\rightarrow t} = \frac{\mathrm{ARP}_1 - \mathrm{ARP}_t}{\mathrm{ARP}_1}.
\]
Higher positive RC indicates larger degradation, lower positive RC indicates better preservation of effectiveness, and negative RC would indicate improvement over snapshot 1. DRI compares a model's relative change against the pivot system, which is the official BM25 baseline in the provided report. A DRI of 0 means the model changes like the pivot; negative DRI means better temporal preservation than the pivot, while positive DRI means worse temporal preservation. The official summary table reports nDCG@10-based longitudinal measures; MAP, Recall@100, and Recall@1000 are therefore treated as development diagnostics unless separately reported by the official evaluation.

To keep the paper readable while preserving exact reproducibility, we treat the submitted identifiers as model names and pair them with explicit structural descriptions. Table~\ref{tab:model-structure-alignment} gives the alignment used in this paper. The model named \texttt{delete-me} in the official report duplicates the BM25 baseline and is excluded from analysis.

\begin{table}[t]
\centering
\small
\caption{Alignment between submitted model names and explicit model structures.}
\label{tab:model-structure-alignment}
\begin{tabular}{p{0.30\linewidth}p{0.38\linewidth}p{0.22\linewidth}}
\toprule
Model name & Explicit structure & Role \\
\midrule
\texttt{baseline-bm25} & BM25 over title+abstract fields & official sparse pivot \\
\texttt{baseline-qwen3-4b} & Qwen3 dense retrieval over title+abstract fields & official dense baseline \\
\texttt{custom-lexical-FT} & BM25 over full text & full-text lexical submission \\
\texttt{bm25-ft-additive} & full-text BM25 with additive score integration & additive integration submission \\
\texttt{bm25-ft-router} & full-text BM25 with router-based integration & router integration submission \\
\texttt{custom-lexical-FT-T} & full-text BM25 with temporal integration & temporal full-text submission \\
\texttt{custom-lexicalft-tc} & full-text BM25 with temporal and citation integration & temporal+citation full-text submission \\
\bottomrule
\end{tabular}
\end{table}

\begin{table}[t]
\centering
\caption{Snapshot-1 month-filtered evaluation views. Documents are filtered by \texttt{publishedDate}; qrels are retained only when the judged document is available in the cumulative document subset.}
\label{tab:monthly-protocol}
\begin{tabular}{lrrrr}
\toprule
Split & Months & Queries & Documents & Purpose \\
\midrule
March & 3 & 74 & 173,031 & early-corpus view \\
March+April & 3,4 & 92 & 343,421 & intermediate growth view \\
March+April+May & 3,4,5 & 96 & 525,293 & full month-filtered view \\
\bottomrule
\end{tabular}
\end{table}

This monthly protocol is an internal simulation rather than an official LongEval test condition. It should be read as a \emph{publication-window diagnostic}: documents are grouped by whether their \texttt{publishedDate} falls inside the March, March--April, or March--May publication window, and qrels are retained only when the judged document is available in that filtered view. We use this rule to study how systems behave on progressively wider publication-date windows inside snapshot-1, not to exactly reconstruct the organizers' later snapshots. A stricter live-corpus simulation would instead include all documents published before or equal to each month-end; we leave that variant to later-snapshot evaluation. The query count changes because queries without at least one retained judgment are excluded from that split. Documents without usable publication dates, documents outside the March--May window, and judged documents not retained by the current cumulative filter are not represented in a given view. This means the monthly results should be interpreted only as a controlled development diagnostic under a shared filtering rule, not as a complete official evaluation or as a direct replacement for later-snapshot testing.

\section{Cascade Model Designs}
\label{sec:model-designs}

We describe the systems as a retrieval cascade, following the common distinction between an initial retrieval stage and later reranking stages \cite{matveeva_nested_ranker_2006}. The first level is the \emph{first-stage retriever}: a lexical, dense, expansion, or run-level fusion method that produces a candidate ranking from an existing index or released run. The second level is the \emph{reranking overlay}: an optional temporal or citation-aware layer applied to candidates from the first stage. This separation is deliberate. It keeps expensive indexing decisions independent from lighter ranking experiments, and it lets us test temporal and citation evidence without changing the underlying document representation.

The suffix \texttt{\_dctr} in local report files means that the run was evaluated with DCTR qrels. It does not mean the model was trained on DCTR labels. The suffixes \texttt{\_temporal} and \texttt{\_citation} identify local reranking variants. To reduce naming ambiguity, the main text uses a fixed set of short names: TA means title+abstract, FT means full text, FT BM25 means BM25 over full text, FT BM25+T means full-text BM25 with temporal integration, and FT BM25+TC means full-text BM25 with temporal and citation integration. Submitted model identifiers are given once in Table~\ref{tab:model-structure-alignment}.

\subsection{First-Stage Retrieval and Run Reuse}

We maintain two lexical document views. The first is title+abstract, which is the compact view used by the official PyTerrier BM25 baseline and the official dense baseline. The second is full text, which is indexed separately for the custom full-text BM25 model. The dense Qwen baseline is not a lexical Terrier index: it uses a dense embedding representation over the same compact textual view. Thus, title+abstract BM25 and dense title+abstract retrieval may consume similar text fields, but they remain different retrieval backends.

All fusion, reranking, temporal, and citation experiments operate within this cascade over first-stage rankings. RRF reads run files; RM3 rewrites the query inside a lexical pipeline but uses the same lexical index; cross-encoder reranking rescores candidate documents; temporal and citation overlays rescore already retrieved candidates. None of these steps requires building a new index. This matters practically because full-text and dense indexes are expensive to construct, while run-level and reranking overlays can be iterated much faster.

\subsection{First-Stage Retrieval Models}

Table~\ref{tab:skeleton-models} summarizes the first-stage systems. These are the main retrieval foundations evaluated before temporal and citation evidence is added.

\begin{table}[t]
\centering
\caption{First-stage retrieval models. TA denotes title+abstract and FT denotes full text.}
\label{tab:skeleton-models}
\begin{tabular}{p{0.18\linewidth}p{0.18\linewidth}p{0.55\linewidth}}
\toprule
Model & Source view & Design role \\
\midrule
TA BM25 & TA & Official PyTerrier lexical anchor. It indexes compact scientific metadata text and provides the pivot run for temporal-change metrics. \\
Official Qwen3 dense & TA & Official dense semantic anchor using Qwen/Qwen3-Embedding-4B. It tests whether neural semantic matching improves over sparse lexical matching. \\
FT BM25 & FT & Custom full-text lexical baseline. It tests whether longer scientific text improves retrieval despite higher indexing cost. \\
TA BM25+RM3 & TA & Query-expansion baseline. It uses an initial BM25 search, selects pseudo-relevant documents, expands the query with feedback terms, and runs BM25 again. \\
TA rerank & TA & Two-stage lexical-neural baseline. It retrieves candidates with TA BM25 and reranks the candidate head using a MiniLM-L-12 MS MARCO cross-encoder. \\
RRF fusion & run files & Run-level fusion over existing ranked lists. It combines complementary sparse, dense, and full-text evidence without score normalization or index rebuilding. \\
\bottomrule
\end{tabular}
\end{table}

The TA BM25 first-stage retriever is the simplest official anchor. It uses the title+abstract representation, Terrier's standard text processing, and BM25 scoring. It is useful not because it is expected to be optimal, but because it provides a stable reference point for both absolute metrics and pivot-relative robustness metrics.

The Official Qwen3 dense first-stage retriever represents a different matching hypothesis. Instead of scoring exact or stemmed term overlap, it encodes queries and documents into an embedding space and retrieves by vector similarity. We use it as a released official dense baseline when comparing results. When reproduced locally, it depends on a local OpenAI-compatible embedding service serving Qwen/Qwen3-Embedding-4B; no external API call is part of the evaluation pipeline.

The FT BM25 first-stage retriever changes the document view rather than the scoring family. It applies BM25 to full text, which can capture evidence absent from abstracts but may also introduce noise from long documents. In the snapshot-1 development results, this model is the strongest single-system baseline; in the official results, the strongest submitted models build on this full-text foundation with temporal integration. Together, these findings suggest that scientific retrieval benefits substantially from full-text lexical evidence.

The TA BM25+RM3 first-stage retriever changes the query rather than the index. RM3 first retrieves documents with BM25, estimates useful feedback terms from the top pseudo-relevant documents, interpolates those terms with the original query, and then runs a second BM25 search. In our configuration, \texttt{fbDocs}=3, \texttt{fbTerms}=10, and \texttt{fbLambda}=0.6. This makes RM3 a non-temporal reformulation method: it aims to reduce vocabulary mismatch, not to exploit future judgments or time signals.

The TA rerank stage changes the second-stage scorer. It keeps BM25 as a candidate generator but rescores top candidates with a cross-encoder that jointly reads the query and candidate document text. This model tests whether neural interaction reranking can improve top-rank precision beyond the initial sparse ranking.

\subsection{Run-Level Fusion}

RRF is treated as a first-stage model family because each fused run becomes a complete non-temporal retrieval system. For each query-document pair, the fused score is
\[
\mathrm{RRF}(d) = \sum_i \frac{1}{k + r_i(d)},
\]
where $r_i(d)$ is the rank of document $d$ in input run $i$. We use $k=60$. If a document is absent from an input run, that run contributes zero. We evaluate three fusion variants:
\begin{itemize}[leftmargin=1.5em]
    \item RRF(TA BM25, Dense): compact lexical plus compact dense evidence.
    \item RRF(FT BM25, Dense): full-text lexical plus compact dense evidence.
    \item RRF(TA BM25, FT BM25, Dense): three-way fusion across compact lexical, full-text lexical, and dense semantic views.
\end{itemize}

RRF is especially useful in this project because BM25, dense retrieval, and reranking scores are not naturally calibrated to each other. Rank-based fusion avoids raw-score normalization and allows us to test whether complementary retrieval views improve robustness. It also cleanly follows the ``overlay'' philosophy: the fused model is built from existing run files, not from a new index.

\subsection{Temporal Overlay}
\label{sec:temporal-overlay}

The temporal overlay is a candidate reranking layer. It takes a first-stage run, loads metadata for the candidate documents, computes time-aware features relative to the active cutoff, and rescales the candidate head. It does not modify the index and does not retrieve new documents. The current implementation uses \texttt{publishedDate} as the primary document-time field, because earlier experiments found that \texttt{updatedDate} and \texttt{createdDate} were too sparse or poorly aligned with the training period.

The overlay first applies a lightweight rule-based query-intent classifier with current/evolving, survey-like, and foundational categories. Its purpose is only to avoid applying the same recency bias to every query.

For a document $d$ and cutoff time $t$, the main temporal features are:
\[
\mathrm{age}(d,t) = t - \mathrm{publishedDate}(d),
\]
\[
\mathrm{recency}(d,t) = \exp(-\mathrm{age}(d,t)/h_f),
\]
\[
\mathrm{foundation}(d,t) = 1 - \exp(-\mathrm{age}(d,t)/h_a).
\]
Here $h_f$ is a freshness half-life and $h_a$ is an age half-life for foundational documents. The overlay also computes a lightweight lexical novelty signal from terms in the candidate document. At a high level, this signal increases when a candidate contains novelty-indicator terms or other document terms that are not already present in the query:
\[
\mathrm{novelty}(q,d) = \lambda_1\mathrm{novelTerms}(q,d) + \lambda_2\mathrm{unseenTerms}(q,d),
\]
where $\mathrm{novelTerms}(q,d)$ summarizes unseen domain-specific novelty indicators and $\mathrm{unseenTerms}(q,d)$ summarizes the broader fraction of document terms not appearing in the query. The implementation clips this combined score to $[0,1]$. The temporal score is then a query-intent-dependent weighted combination of recency, foundation, update, and novelty features:
\[
T(q,d,t) = w_r(q)\mathrm{recency}(d,t) + w_f(q)\mathrm{foundation}(d,t) + w_n(q)\mathrm{novelty}(q,d).
\]
For current or evolving queries, the overlay gives more weight to recency and novelty. For foundational or survey-like queries, it gives more weight to age-stable evidence and less weight to recency. The local overlay should be read as a heuristic first pass: it reranks only the candidate head, keeps the retrieved tail available, and uses query-level score normalization before interpolation with the base score. Because the snapshot-1 diagnostic experiments below show very large drops for some sparse local overlays, we treat those drops as evidence of a calibration or weighting problem rather than as evidence that temporal metadata is intrinsically unhelpful.

\subsection{Citation Overlay}
\label{sec:citation-overlay}

The citation overlay uses a local OpenCitations-derived graph to compute candidate-level features such as inbound citations, recent inbound citations, citation velocity, and outbound citations. We derive two summary signals. The foundational signal rewards documents with many inbound citations but a low recent-citation ratio:
\[
C_{\mathrm{foundation}}(d,t) = \log(1+\mathrm{inbound}(d,t))\cdot(1-\mathrm{recentRatio}(d,t)).
\]
The emerging citation signal rewards documents receiving recent citation attention:
\[
C_{\mathrm{emerging}}(d,t) = \log(1+\mathrm{recentInbound}(d,t))\cdot \mathrm{recentRatio}(d,t).
\]
In these equations, \(\mathrm{inbound}(d,t)\) is the number of citations received by document \(d\) up to cutoff \(t\), \(\mathrm{recentInbound}(d,t)\) is the subset of those received citations whose edge timestamp falls in the recent window before \(t\), \(\mathrm{outbound}(d,t)\) denotes references made by \(d\) that are visible by \(t\), and \(\mathrm{recentRatio}(d,t)=\mathrm{recentInbound}(d,t)/\max(1,\mathrm{inbound}(d,t))\). We use the more familiar terms \emph{citations} and \emph{references} in the prose, but keep inbound/outbound notation in the formulas to make graph direction explicit.

Citation features are computed as of the active cutoff. Citation edges after the cutoff are removed when timestamp information is available. Candidate documents without citation metadata receive zero-valued citation features rather than being dropped. This keeps the overlay safe for sparse citation coverage, but it also means citation features can be weak when the graph is incomplete. Because we do not yet report a full citation-coverage audit, the citation results should be interpreted conservatively: a weak overlay score may reflect sparse or timestamp-incomplete graph coverage as much as the intrinsic value of citation evidence.

\subsection{Complete Overlay Models}

The complete overlay score combines the normalized first-stage score, temporal score, and citation score:
\[
S'(q,d,t) = \alpha \widehat{S}_{\mathrm{base}}(q,d)
          + \beta(q) \widehat{T}(q,d,t)
          + \gamma(q) \widehat{C}(q,d,t).
\]
The temporal-only model sets $\gamma(q)=0$. A citation-only variant would set $\beta(q)=0$, while the temporal+citation model allows both terms. In the current completed experiments, we evaluate temporal-only overlays for several first-stage runs and temporal+citation overlays for selected first-stage runs. The citation-only ablation remains useful but is not yet the primary reported setting.

The official submitted full-text variants instantiate this idea in three ways. FT BM25+additive and FT BM25+router are integration controls over the full-text lexical foundation, while FT BM25+temporal and FT BM25+temporal+citation are the submitted temporal and temporal+citation systems. The official results therefore let us ask whether temporal and citation evidence help under the shared external evaluation, while the local overlay tables below explain which diagnostic configurations remain unstable.

Table~\ref{tab:overlay-models} lists how the first-stage and reranking families relate to each other. The important point is that a temporal or citation model is not a new first-stage retriever. It is a reranked version of a first-stage model, and therefore it can be compared directly against the original first-stage run to measure the value or harm of the overlay.

\begin{table}[t]
\centering
\caption{Relationship between first-stage models and overlay variants.}
\label{tab:overlay-models}
\small
\begin{tabular}{p{0.22\linewidth}p{0.14\linewidth}p{0.14\linewidth}p{0.32\linewidth}}
\toprule
Variant type & Temporal reranker & Citation reranker & Interpretation \\
\midrule
First-stage run & no & no & Original retrieval or fusion run. \\
Temporal-only & yes & no & Tests whether publication-time features improve candidate order. \\
Citation-only & no & yes & Planned clean ablation for citation graph evidence alone. \\
Temporal+citation & yes & yes & Tests whether citation evidence helps after temporal reranking. \\
\bottomrule
\end{tabular}
\normalsize
\end{table}

\section{Results}

\subsection{Official Three-Snapshot Evaluation}

Table~\ref{tab:official-results} reports the official DCTR results provided for Task 1. The most important change relative to the development-only story is that the temporalized full-text submissions are strongest in the official evaluation. FT BM25+temporal and FT BM25+temporal+citation obtain the best ARP on all three snapshots: 0.285 on snapshot 1, 0.267 on snapshot 2, and 0.180 on snapshot 3. They also reduce RC on snapshot 3 from 0.481 for the Official BM25 pivot to 0.368. In this table, lower RC means less degradation from snapshot 1, and negative DRI values indicate better preservation than the BM25 pivot under the official pivot-relative calculation.

\begin{table}[t]
\centering
\scriptsize
\caption{Official Task 1 DCTR results. ARP, RC, and DRI are reported for nDCG@10.}
\label{tab:official-results}
\begin{tabular}{p{0.22\linewidth}p{0.18\linewidth}rrrrrrr}
\toprule
Model name & Explicit structure & \multicolumn{3}{c}{ARP} & \multicolumn{2}{c}{RC} & \multicolumn{2}{c}{DRI} \\
\cmidrule(lr){3-5}\cmidrule(lr){6-7}\cmidrule(lr){8-9}
 & & S1 & S2 & S3 & S2 & S3 & S2 & S3 \\
\midrule
\texttt{baseline-bm25} & BM25 over TA & 0.270 & 0.236 & 0.140 & 0.127 & 0.481 & 0.000 & 0.000 \\
\texttt{baseline-qwen3-4b} & Qwen3 dense over TA & 0.228 & 0.189 & 0.125 & 0.171 & 0.452 & 0.042 & -0.047 \\
\texttt{custom-lexical-FT} & BM25 over FT & 0.275 & 0.239 & 0.153 & 0.131 & 0.445 & 0.005 & -0.071 \\
\texttt{bm25-ft-additive} & FT BM25, additive integration & 0.276 & 0.240 & 0.153 & 0.128 & 0.446 & 0.001 & -0.070 \\
\texttt{bm25-ft-router} & FT BM25, router integration & 0.276 & 0.240 & 0.153 & 0.131 & 0.446 & 0.004 & -0.070 \\
\texttt{custom-lexical-FT-T} & FT BM25, temporal integration & 0.285 & 0.267 & 0.180 & 0.064 & 0.368 & -0.077 & -0.230 \\
\texttt{custom-lexicalft-tc} & FT BM25, temporal+citations & 0.285 & 0.267 & 0.180 & 0.064 & 0.368 & -0.077 & -0.230 \\
\bottomrule
\end{tabular}
\end{table}

The official comparison supports three conclusions. First, full-text lexical retrieval improves over both official baselines, especially by snapshot 3. Second, the temporal full-text variants are the best official submissions and are more robust than the BM25 pivot according to both RC and DRI. Third, the temporal+citation submission matches the temporal-only submission exactly in the official summary, so citation evidence is not harmful here but its added value is not demonstrated by these aggregate results.

\subsection{Snapshot-1 Development Effectiveness}

Table~\ref{tab:whole-train} reports the strongest DCTR whole-train development results on snapshot-1 train. These results are not a replacement for the official evaluation in Table~\ref{tab:official-results}; they explain the local model-selection process and include deeper metrics not shown in the official summary. Full-text BM25 is the best single development model on nDCG@10 and MAP. The RRF run combining full-text BM25 with dense title+abstract retrieval has slightly lower nDCG@10 but higher Recall@100 and Recall@1000, making it a strong candidate-generation strategy.

\begin{table}[t]
\centering
\caption{Snapshot-1 train DCTR results.}
\label{tab:whole-train}
\begin{tabular}{lrrrrr}
\toprule
Method & nDCG@10 & nDCG@1000 & MAP & R@100 & R@1000 \\
\midrule
FT BM25 & 0.3302 & 0.5077 & 0.2853 & 0.7394 & 0.9245 \\
TA rerank & 0.3222 & 0.4821 & 0.2777 & 0.6952 & 0.8581 \\
RRF(FT,Dense) & 0.3175 & 0.5173 & 0.2749 & 0.7806 & 0.9667 \\
RRF(TA,FT,Dense) & 0.3028 & 0.5084 & 0.2572 & 0.7870 & 0.9806 \\
TA BM25 & 0.2922 & 0.4564 & 0.2573 & 0.6836 & 0.8581 \\
Official Qwen3 dense & 0.2820 & 0.4483 & 0.2378 & 0.6390 & 0.8613 \\
TA BM25+RM3 & 0.2781 & 0.4510 & 0.2402 & 0.6559 & 0.8701 \\
\bottomrule
\end{tabular}
\end{table}

The same ordering is broadly consistent under raw qrels: FT BM25 remains first on nDCG@10 (0.3637), while RRF(FT,Dense) remains close behind (0.3532) and has the strongest deep recall among the highlighted runs. We report these raw-qrel figures only as a sensitivity check in the current draft; DCTR remains the primary qrel variant used for the main tables.

\subsection{Local Temporal and Citation Overlay Diagnostics}

Table~\ref{tab:temporal-results} reports first-pass temporal and citation overlays in the local snapshot-1 development setting. These diagnostics are mostly negative even though the official temporal full-text submissions in Table~\ref{tab:official-results} are positive. The temporal overlay improves the official dense baseline over dense alone (0.3069 vs. 0.2820 nDCG@10), but it does not beat full-text BM25 or the best RRF fusion run in the local development evaluation. On sparse full-text and title+abstract BM25, the simple local temporal scoring severely hurts top-rank quality. Citation features also do not recover the loss in this diagnostic configuration.

\begin{table}[t]
\centering
\caption{Temporal and citation overlays on snapshot-1 train DCTR.}
\label{tab:temporal-results}
\begin{tabular}{lrrrrr}
\toprule
Method & nDCG@10 & nDCG@1000 & MAP & R@100 & R@1000 \\
\midrule
Local Dense+temporal & 0.3069 & 0.4629 & 0.2551 & 0.6400 & 0.8613 \\
Local TA rerank+temporal & 0.2680 & 0.4307 & 0.2354 & 0.5447 & 0.8581 \\
Local TA rerank+temp+citation & 0.2543 & 0.4194 & 0.2220 & 0.5451 & 0.8581 \\
Local TA RM3+temporal & 0.0154 & 0.1842 & 0.0144 & 0.0433 & 0.8701 \\
Local FT BM25+temporal & 0.0088 & 0.1935 & 0.0091 & 0.0213 & 0.9245 \\
Local TA BM25+temporal & 0.0053 & 0.1750 & 0.0084 & 0.0298 & 0.8581 \\
\bottomrule
\end{tabular}
\end{table}

This failure mode is informative. The overlay preserves tail recall because only the head is reranked, but it can move weakly relevant temporally favored documents into top ranks. The contrast with the official results suggests that temporal evidence can help when applied to the full-text submission pipeline, but that local overlay implementations remain sensitive to score normalization, candidate-head choice, and feature weights. The large sparse-system drops should therefore be read as a calibration/debugging signal: future versions need no-op sanity checks, per-query score inspection, and separate temporal-only and citation-only calibration before combining signals.

\subsection{Cumulative Monthly Diagnostic Evaluation}

Table~\ref{tab:monthly-results} reports selected cumulative monthly DCTR results from the internal snapshot-1 diagnostic protocol. The best model across all three monthly splits is \texttt{rrf\_bm25\_ft\_dense\_ta}. It also has the smallest relative change from March+April to March+April+May among strong systems, which means it is already strong early and remains strong as the corpus grows. These monthly results are useful for explaining model behavior, but the official three-snapshot results in Table~\ref{tab:official-results} remain the primary longitudinal evidence.

\begin{table}[t]
\centering
\caption{Cumulative monthly DCTR nDCG@10.}
\label{tab:monthly-results}
\begin{tabular}{lrrr}
\toprule
Method & March & March+April & March+April+May \\
\midrule
RRF(FT,Dense) & 0.2732 & 0.2849 & 0.2930 \\
Local Dense+temporal & 0.2241 & 0.2792 & 0.2917 \\
FT BM25 & 0.2314 & 0.2259 & 0.2495 \\
TA rerank & 0.1728 & 0.2172 & 0.2655 \\
TA BM25 & 0.1767 & 0.2047 & 0.2377 \\
Official Qwen3 dense & 0.2009 & 0.2182 & 0.2316 \\
\bottomrule
\end{tabular}
\end{table}

\subsection{Internal Pivot-Relative Temporal Change Metrics}

The monthly curves show absolute performance, but LongEval-style analysis also asks how much a system changes relative to a pivot. We therefore compute five temporal-change metrics from the monthly split report using \texttt{official\_pyterrier\_dctr} as the pivot and nDCG@10 as the primary score. For a transition from earlier split $a$ to later split $b$, ARP is the later score, MARP is the mean of the earlier and later scores, and relative change uses the same sign convention as the official RC metric:
\[
\mathrm{RC} = \frac{s_a - s_b}{s_a}.
\]
Negative RC means that the system improved from the earlier split to the later split; positive RC means degradation. DRI subtracts the pivot RC from the system RC, and ER is the ratio between system RC and pivot RC. These metrics should be read with ARP and MARP rather than alone, because a low-scoring system can show a large relative change from a very small baseline.

\begin{table}[t]
\centering
\caption{Temporal-change metrics for the March to March+April+May transition, using nDCG@10 and TA BM25 as the pivot.}
\label{tab:temporal-change}
\begin{tabular}{lrrrrr}
\toprule
Method & ARP & MARP & RC & DRI & ER \\
\midrule
RRF(TA,FT,Dense) & 0.2526 & 0.2477 & -0.0408 & 0.3041 & 0.1182 \\
RRF(FT,Dense) & 0.2930 & 0.2831 & -0.0725 & 0.2724 & 0.2102 \\
FT BM25 & 0.2495 & 0.2405 & -0.0784 & 0.2665 & 0.2273 \\
Official Qwen3 dense & 0.2316 & 0.2162 & -0.1531 & 0.1917 & 0.4440 \\
Local Dense+temporal & 0.2917 & 0.2579 & -0.3017 & 0.0432 & 0.8749 \\
TA BM25 pivot & 0.2377 & 0.2072 & -0.3449 & 0.0000 & 1.0000 \\
TA rerank & 0.2655 & 0.2192 & -0.5364 & -0.1915 & 1.5552 \\
Local TA rerank+temp+citation & 0.2165 & 0.1871 & -0.3734 & -0.0285 & 1.0828 \\
\bottomrule
\end{tabular}
\end{table}

The most informative comparison is between high absolute performance and relative movement. RRF(FT,Dense) has the highest ARP in Table~\ref{tab:temporal-change}, while also having much smaller relative movement than the pivot (ER = 0.2102). RRF(TA,FT,Dense) has the smallest relative movement among the selected systems (ER = 0.1182), but its ARP is lower than RRF(FT,Dense). This distinction is exactly why we report both absolute scores and change metrics: a system can be robust because it barely moves, or useful because it remains both high-scoring and stable.

The local temporal overlay on Official Qwen3 dense is also instructive. It has strong ARP and MARP under the monthly protocol, but its RC is closer to the pivot, meaning it changes more across the full March-to-May transition than the RRF systems. Meanwhile, some local temporal/citation overlays have poor absolute performance despite moderate or large change values. These results suggest that temporal-change metrics are valuable diagnostics, but they cannot replace direct retrieval effectiveness metrics.

\section{Qualitative System Update-Monitoring Triggers}
\label{sec:update-triggers}

Beyond ranking effectiveness, Task 1 raises a maintenance question: when should a retrieval system be updated as the scientific collection continues to grow? This section should be read as a qualitative monitoring analysis, not as a quantitative backtest of update benefit: for all 100 training queries with dated relevant documents, the latest judged-relevant document already appears before 2025-03-01, so reindexing during March--May mostly adds later distractors rather than newly judged-relevant answers. Our current answer is deliberately simple. Rather than training a learned controller, we define a weekly monitoring policy over collection-drift proxies and interpret an update as an incremental refresh of the retrieval state that can be applied without redesigning the ranking stack from Section~\ref{sec:model-designs}.

\subsection{Weekly Monitoring Policy}

The selected policy has two practical trigger branches. The first is burst detection through \texttt{velocity\_per\_day}, which measures weekly document-ingestion velocity and fires when the current week is much faster than the recent baseline. The second is structural drift detection through the conjunction of \texttt{staleness\_rate} and \texttt{coverage\_gap}. Intuitively, \texttt{staleness\_rate} measures how much of the indexed collection has become old under the chosen stale-age rule: a stale document is one whose publication date is older than the selected stale-age threshold relative to the live collection date used for that monitoring week. \texttt{coverage\_gap} measures how far the indexed anchor has drifted from the live collection size. We also track \texttt{temporal\_gap\_growth\_days}, which measures how much the collection's time-lag profile has worsened relative to the baseline state, but in the final selected strategy it is not the decisive trigger.

This metric split is intentional. \texttt{velocity\_per\_day} is meant to catch sudden ingestion shocks, while \texttt{staleness\_rate} + \texttt{coverage\_gap} is meant to catch slower structural drift. In other words, the policy updates either when the collection changes unusually fast or when the indexed view has become both too old and too incomplete.

\begin{table}[t]
\centering
\scriptsize
\caption{Quantitative design of the monitoring metrics used by the update policy.}
\label{tab:update-metrics}
\begin{tabular}{@{}p{0.23\linewidth}p{0.27\linewidth}p{0.22\linewidth}p{0.16\linewidth}@{}}
\toprule
Metric & Definition & High value indicates & Role \\
\midrule
\texttt{velocity\_per\_day} & weekly new documents divided by $7$ & an ingestion burst & primary trigger \\
\texttt{staleness\_rate} & stale indexed documents divided by indexed documents & the indexed mix is old & structural trigger \\
\texttt{coverage\_gap} & $(N_{\mathrm{live}} - N_{\mathrm{anchor}}) / N_{\mathrm{anchor}}$ & the index lags the live corpus & structural trigger \\
\texttt{temporal gap growth} & current temporal gap minus baseline temporal gap in days & time lag is worsening & diagnostic only \\
\bottomrule
\end{tabular}
\end{table}

Table~\ref{tab:update-metrics} makes the metric design explicit. \(N_{\mathrm{live}}\) is the number of documents in the live collection at the monitoring time, and \(N_{\mathrm{anchor}}\) is the document count in the indexed anchor state being monitored. \texttt{velocity\_per\_day} is the most reactive signal because it directly measures current ingestion pressure. \texttt{staleness\_rate} and \texttt{coverage\_gap} are less about one-week spikes and more about accumulated mismatch between the deployed retrieval state and the live collection. \texttt{temporal\_gap\_growth\_days} is still useful descriptively, but it is weaker as a practical decision signal in the current setup.

Our final parameterization keeps this policy simple. We estimate the baseline velocity from the first 4 weeks, trigger the burst branch when weekly velocity exceeds $3.0\times$ that baseline, trigger the structural branch when \texttt{staleness\_rate} reaches 0.80 and \texttt{coverage\_gap} reaches 0.03, and keep \texttt{temporal\_gap\_growth\_days} = 20 as a diagnostic threshold. In shorthand, the update rule is
\[
\text{update}_t = \mathbf{1}[v_t > 3\bar{v}_{1:4}]\;\vee\;\mathbf{1}[s_t \ge 0.80 \wedge c_t \ge 0.03],
\]
where $v_t$ is weekly ingestion velocity, $\bar{v}_{1:4}$ is the baseline velocity estimated from the first four weeks, $s_t$ is the staleness rate, and $c_t$ is the coverage gap. Operationally, this means that the system monitors the collection week by week and issues an incremental update when either the burst branch or the structural-drift branch becomes active. In the final selected timeline, most updates come from the burst branch, so the policy behaves primarily as a velocity-sensitive monitoring rule with a stale-coverage fallback condition.

\subsection{Why We Treat This as a Qualitative Analysis}

As noted at the start of this section, we interpret the policy as scheduling/monitoring analysis, not as a causal claim that more frequent updates improve current-qrel effectiveness.

\subsection{Selected Timeline and Interpretation}

Under the final weekly strategy, updates are concentrated in the middle of the year: 2 in snapshot-1, 5 in snapshot-2, and 1 in snapshot-3. This suggests bursty ingestion pressure rather than smooth temporal lag. The first seven updates are triggered by ingestion velocity; only the final late-period update also reflects stale-coverage drift.

\begin{figure}[t]
\centering
\includegraphics[width=0.92\linewidth]{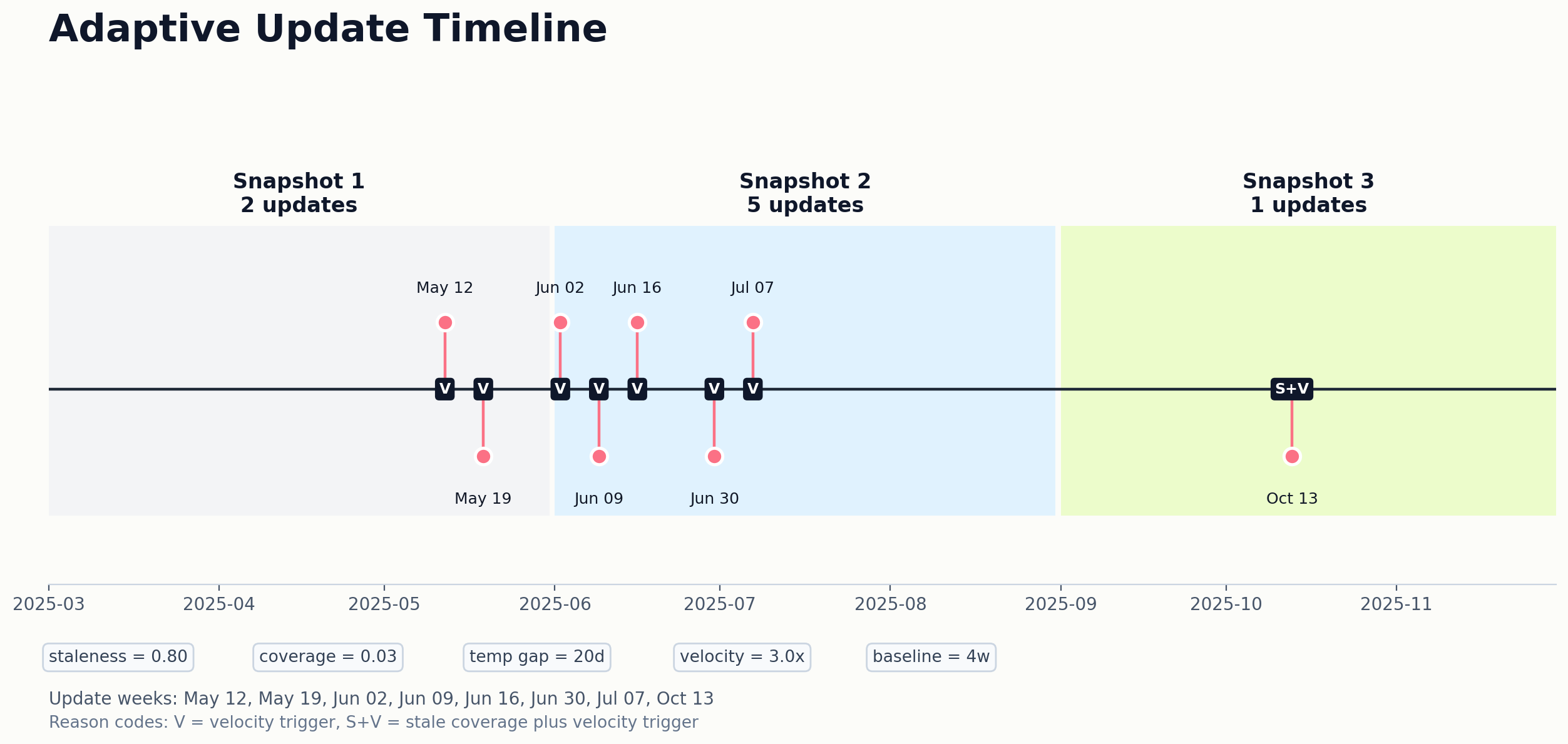}
\caption{Selected weekly update timeline for the final policy. The update pattern is concentrated in the middle period, yielding a 2/5/1 distribution across the three snapshot-aligned windows.}
\label{fig:update-timeline}
\end{figure}

\section{Discussion}

The main practical lesson is that strong full-text lexical retrieval is the foundation for our best LongEval-Sci systems, and that temporal integration can improve that foundation in the official longitudinal evaluation. FT BM25 already improves over the official BM25 and Qwen3 dense baselines, while FT BM25+temporal and FT BM25+temporal+citation achieve the strongest official ARP on all three snapshots and reduce snapshot-3 degradation relative to the BM25 pivot. This changes the interpretation from ``temporal evidence failed'' to ``temporal evidence helps in the submitted full-text pipeline, but remains easy to miscalibrate in local overlays.''

Field choice matters strongly. Full-text BM25 is the best single-system anchor in snapshot-1 development and remains competitive in the official evaluation. This suggests that, for LongEval-Sci, important relevance evidence often appears outside the abstract. At the same time, full text is not uniformly superior for every role: title+abstract remains attractive for compact indexing, dense retrieval, and reranking inputs because it reduces document length and noise.

Fusion and reranking play different roles in the development analysis. Cross-encoder reranking improves over the official TA BM25 baseline on the whole-train setting, but RRF is more attractive as a robust candidate-generation strategy. RRF(FT,Dense) does not beat FT BM25 on snapshot-1 nDCG@10, but it substantially improves Recall@1000 and performs best under the internal cumulative monthly nDCG@10 protocol. This makes it a strong candidate for future official submissions, because a reranker cannot promote relevant documents that never appear in the candidate pool.

RM3 is weaker in the current configuration. It may suffer from short queries, noisy feedback documents, or untuned parameters; this is not evidence against query expansion generally, only against this setting relative to full-text retrieval and fusion.

The temporal and citation evidence should be interpreted in two layers. In the official evaluation, temporal full-text retrieval is clearly helpful: it improves ARP and robustness over FT BM25 and the BM25 pivot. In the local diagnostic evaluation, however, some temporal overlays over sparse systems sharply hurt nDCG@10. This indicates that temporal features can overpower the base relevance score if normalization or weights are wrong. Citation evidence is more tentative: FT BM25+temporal+citation exactly matches FT BM25+temporal in the official summary, so citation is not harmful in that submitted model, but its incremental value is not demonstrated.

The monthly protocol is useful as a development diagnostic but should not be overclaimed. It filters documents and qrels by \texttt{publishedDate}, so it simulates cumulative corpus growth inside snapshot-1 rather than the official snapshot-1/2/3 evaluation. Query counts also vary across splits because queries without retained judgments are excluded. For that reason, the monthly results are best used for model selection and robustness stress-testing, while the official results in Table~\ref{tab:official-results} should drive the final longitudinal conclusions.

For participants building LongEval-Sci systems, these results suggest three practical lessons. First, full-text lexical retrieval is a strong and inexpensive improvement over compact title+abstract baselines when full text is available. Second, temporal evidence can improve full-text retrieval under the official evaluation, but should be integrated with calibration checks rather than as a uniform boost. Third, RRF is a promising candidate-generation overlay because it improves deep recall without score calibration or index rebuilding, even though it was evaluated here mainly as an internal development diagnostic.

The update-monitoring analysis adds a complementary systems perspective to these ranking results. In the current data setting, the proposed weekly policy is driven mostly by ingestion bursts and secondarily by stale-coverage drift, not by the arrival of newly judged-relevant documents. This matters for interpretation: update decisions are operational responses to collection change, whereas ranking comparisons remain the right tool for deciding which retrieval foundation to deploy within a given index state.

\section{Limitations}

This research has several limitations. The official summary available to us reports nDCG@10-based ARP, RC, and DRI, while MAP and recall are reported only for our internal development diagnostics. The current paper also does not yet package the complete software and run artifacts for public reuse, which limits reproduction beyond the descriptions and tables reported here. The temporal+citation official model matches the temporal-only model exactly in the aggregate table, so the added value of citation evidence is not established. The large sparse-overlay drops in local diagnostics require no-op reproduction checks, score inspection, and per-query error analysis. Citation timestamp coverage and unknown-timestamp handling need a deeper audit. We do not yet report paired significance tests. The monthly split may be affected by judgment incompleteness, publication-date missingness, and changing query counts.

\section{Future Work}

Future work should focus on explaining and strengthening the official temporal full-text result. This includes citation-only and temporal-only ablations under the same official-style protocol, per-query analysis by temporal intent, raw-vs-DCTR sensitivity tables, paired significance tests, and deeper inspection of why the official temporal full-text models succeed while some local sparse overlays fail. We also need a clearer comparison of additive and router variants, stronger model-name consistency between code artifacts and paper tables, public packaging of run files and evaluation scripts, and validation of whether velocity/drift triggers predict downstream effectiveness changes.

\section{Conclusion}

We presented official and diagnostic LongEval-Sci analyses over sparse retrieval, dense retrieval, RM3, reranking, RRF, and temporal/citation integration. The official results show that full-text temporal retrieval is our strongest submitted strategy: FT BM25+temporal and FT BM25+temporal+citation achieve the best ARP on all three snapshots and reduce degradation relative to the BM25 pivot. The internal diagnostics explain the foundation for this result: full-text BM25 is the strongest snapshot-1 single retriever, RRF is the strongest deep-recall candidate-generation strategy, and uncalibrated overlays can still fail badly. Citation evidence remains inconclusive because the temporal+citation model matches temporal-only performance. Finally, our qualitative update-monitoring analysis suggests that refresh decisions should be handled as a separate maintenance layer guided by ingestion velocity and stale-coverage drift.

\section*{Declaration of Generative AI and AI-assisted technologies}

During the preparation of this work, the author(s) used ChatGPT, Grammarly in order to: Grammar and spelling check, Paraphrase and reword. After using this tool/service, the author(s) reviewed and edited the content as needed and take(s) full responsibility for the publication’s content.

\bibliography{longeval2026_yyd}

@inproceedings{dai_learning_2011,
	address = {Beijing China},
	title = {Learning to rank for freshness and relevance},
	isbn = {978-1-4503-0757-4},
	url = {https://dl.acm.org/doi/10.1145/2009916.2009933},
	doi = {10.1145/2009916.2009933},
	language = {en},
	urldate = {2026-04-18},
	booktitle = {Proceedings of the 34th international {ACM} {SIGIR} conference on {Research} and development in {Information} {Retrieval}},
	publisher = {ACM},
	author = {Dai, Na and Shokouhi, Milad and Davison, Brian D.},
	month = jul,
	year = {2011},
	pages = {95--104},
}

@article{zobel_inverted_2006,
	title = {Inverted files for text search engines},
	volume = {38},
	issn = {0360-0300, 1557-7341},
	url = {https://dl.acm.org/doi/10.1145/1132956.1132959},
	doi = {10.1145/1132956.1132959},
	language = {en},
	number = {2},
	urldate = {2026-04-17},
	journal = {ACM Computing Surveys},
	author = {Zobel, Justin and Moffat, Alistair},
	month = jul,
	year = {2006},
	pages = {6},
}

@misc{asadi_fast_2013,
	title = {Fast, {Incremental} {Inverted} {Indexing} in {Main} {Memory} for {Web}-{Scale} {Collections}},
	url = {http://arxiv.org/abs/1305.0699},
	doi = {10.48550/arXiv.1305.0699},
	urldate = {2026-04-17},
	publisher = {arXiv},
	author = {Asadi, Nima and Lin, Jimmy},
	month = may,
	year = {2013},
	note = {arXiv:1305.0699 [cs]},
}

@article{moffat_efficient_2023,
	title = {Efficient {Immediate}-{Access} {Dynamic} {Indexing}},
	volume = {60},
	issn = {03064573},
	url = {http://arxiv.org/abs/2211.06030},
	doi = {10.1016/j.ipm.2022.103248},
	number = {3},
	urldate = {2026-04-17},
	journal = {Information Processing \& Management},
	author = {Moffat, Alistair and Mackenzie, Joel},
	month = may,
	year = {2023},
	note = {arXiv:2211.06030 [cs]},
	pages = {103248},
}

@misc{mohoney_incremental_2024,
	title = {Incremental {IVF} {Index} {Maintenance} for {Streaming} {Vector} {Search}},
	url = {http://arxiv.org/abs/2411.00970},
	doi = {10.48550/arXiv.2411.00970},
	urldate = {2026-04-17},
	publisher = {arXiv},
	author = {Mohoney, Jason and Pacaci, Anil and Chowdhury, Shihabur Rahman and Minhas, Umar Farooq and Pound, Jeffery and Renggli, Cedric and Reyhani, Nima and Ilyas, Ihab F. and Rekatsinas, Theodoros and Venkataraman, Shivaram},
	month = nov,
	year = {2024},
	note = {arXiv:2411.00970 [cs]},
}

@inproceedings{brown_fast_nodate,
	title = {Fast {Incremental} {Indexing} for {Full}-{Text} {Information} {Retrieval}},
	language = {en},
	author = {Brown, Eric W and Callan, James P and Croft, W Bruce},
	booktitle = {Proceedings of the 20th International Conference on Very Large Data Bases (VLDB)},
	publisher = {Morgan Kaufmann},
	pages = {192--202},
	year = {1994},
}

@inproceedings{cormack2009rrf,
	title = {Reciprocal rank fusion outperforms Condorcet and individual rank learning methods},
	booktitle = {Proceedings of the 32nd International ACM SIGIR Conference on Research and Development in Information Retrieval},
	author = {Cormack, Gordon V. and Clarke, Charles L. A. and Buettcher, Stefan},
	year = {2009},
	pages = {758--759},
	doi = {10.1145/1571941.1572114},
}

@inproceedings{lavrenko2001relevance,
	title = {Relevance-based language models},
	booktitle = {Proceedings of the 24th Annual International ACM SIGIR Conference on Research and Development in Information Retrieval},
	author = {Lavrenko, Victor and Croft, W. Bruce},
	year = {2001},
	pages = {120--127},
	doi = {10.1145/383952.383972},
}

@inproceedings{alexander2024openwebsearch,
	title = {Team OpenWebSearch at CLEF 2024: LongEval},
	booktitle = {Working Notes of CLEF 2024},
	author = {Alexander, Daria and Fr{\"o}be, Maik and Hendriksen, Gijs and Schlatt, Ferdinand and Hagen, Matthias and Hiemstra, Djoerd and Potthast, Martin and de Vries, Arjen P.},
	year = {2024},
	pages = {1--10},
}

@inproceedings{alexander2025openwebsearch,
	title = {Team OpenWebSearch at LongEval: Using Historical Data for Scientific Search},
	booktitle = {Working Notes of the Conference and Labs of the Evaluation Forum, CLEF 2025},
	author = {Alexander, Daria and Fr{\"o}be, Maik and Hendriksen, Gijs and Hagen, Matthias and Hiemstra, Djoerd and Potthast, Martin and de Vries, Arjen P.},
	year = {2025},
	series = {CEUR Workshop Proceedings},
	volume = {4038},
	pages = {3316--3326},
}

@inproceedings{agh2025,
	title = {AGH IR at LongEval: Improving Scientific Information Retrieval with Dense Representations and Cross-Encoder Re-ranking},
	booktitle = {Working Notes of CLEF 2025},
	author = {Stryszewski, Jan and Prosowicz, Wiktor and Kawiak, Tomasz and Ja{\'s}kowiec, Adrian},
	year = {2025},
	pages = {3495--3503},
}

@inproceedings{cancellieri_longeval_2025_clean,
	title = {{LongEval} at {CLEF} 2025: {Longitudinal} {Evaluation} of {IR} {Model} {Performance}},
	author = {Cancellieri, Matteo and El-Ebshihy, Alaa and Fink, Tobias and Galu{\v{s}}{\v{c}}{\'a}kov{\'a}, Petra and Gonzalez-Saez, Gabriela and Goeuriot, Lorraine and Iommi, David and Keller, J{\"u}ri and Knoth, Petr and Mulhem, Philippe and Piroi, Florina and Pride, David and Schaer, Philipp},
	editor = {Hauff, Claudia and Macdonald, Craig and Jannach, Dietmar and Kazai, Gabriella and Nardini, Franco Maria and Pinelli, Fabio and Silvestri, Fabrizio and Tonellotto, Nicola},
	booktitle = {Advances in Information Retrieval},
	publisher = {Springer Nature Switzerland},
	address = {Cham},
	year = {2025},
	pages = {382--388},
	isbn = {978-3-031-88720-8},
}

@inproceedings{gade_tempralm_2025,
	title = {It's {About} {Time}: {Incorporating} {Temporality} in {Retrieval} {Augmented} {Language} {Models}},
	url = {https://doi.org/10.1109/CAI64502.2025.00019},
	doi = {10.1109/CAI64502.2025.00019},
	booktitle = {Proceedings of the 2025 {IEEE} Conference on Artificial Intelligence ({CAI})},
	author = {Gade, Anoushka and Jetcheva, Jorjeta G. and Trivedi, Hardi},
	year = {2025},
	pages = {75--82},
}

@article{janssens2015cocites,
	title = {Novel citation-based search method for scientific literature: application to meta-analyses},
	journal = {BMC Medical Research Methodology},
	author = {Janssens, A. Cecile J. W. and Gwinn, Marta},
	year = {2015},
	volume = {15},
	pages = {84},
	doi = {10.1186/s12874-015-0077-z},
}

@article{janssens2020validation,
	title = {Novel citation-based search method for scientific literature: a validation study},
	journal = {BMC Medical Research Methodology},
	author = {Janssens, A. Cecile J. W. and Gwinn, Marta and Brockman, J. Elaine and Powell, Kimberley and Goodman, Michael},
	year = {2020},
	volume = {20},
	pages = {25},
	doi = {10.1186/s12874-020-0907-5},
}

@article{belter2017ranking,
	title = {A relevance ranking method for citation-based search results},
	journal = {Scientometrics},
	author = {Belter, Christopher W.},
	year = {2017},
	volume = {112},
	number = {2},
	pages = {731--746},
	doi = {10.1007/s11192-017-2406-y},
}

@article{bascur2023citationclusters,
	title = {Academic information retrieval using citation clusters: in-depth evaluation based on systematic reviews},
	journal = {Scientometrics},
	author = {Bascur, Juan Pablo and Verberne, Suzan and van Eck, Nees Jan and Waltman, Ludo},
	year = {2023},
	volume = {128},
	pages = {2895--2921},
	doi = {10.1007/s11192-023-04683-1},
}

@inproceedings{lncs_overview_longeval_2026,
  author       = {Timo Breuer and
                  Matteo Cancellieri and
                  Alaa El{-}Ebshihy and
                  Maik Fr{\"o}be and
                  Petra Galusc{\'{a}}kov{\'{a}} and
                  Lorraine Goeuriot and
                  Gabriel Iturra{-}Bocaz and
                  J{\"u}ri Keller and
                  Petr Knoth and
                  Andreas Konstantin Kruff and
                  Philippe Mulhem and
                  Florina Piroi and
                  David Pride and
                  Philipp Schaer and
                  Didier Schwab},
  title        = {Overview of the CLEF 2026 LongEval Lab on Longitudinal Evaluation of Model Performance},
  booktitle    = {Experimental {IR} Meets Multilinguality, Multimodality, and Interaction. Proceedings of the Seventeenth International Conference of the {CLEF} Association ({CLEF} 2026)},
  editor       = {Hagen, Matthias and Potthast, Martin and Stein, Benno and Schaer, Philipp and Zangerle, Eva and MacAvaney, Sean and Stru{\ss}, Julia Maria and S\'{a}nchez Salido, Eva and Barr\'{o}n-Cede\~{n}o, Alberto and Garc\'{i}a Seco de Herrera, Alba},
  publisher    = {Springer},
  series       = {Lecture Notes in Computer Science (LNCS)},
  year         = {2026},
  pages        = {to appear},
}

@inproceedings{matveeva_nested_ranker_2006,
	title = {High Accuracy Retrieval with Multiple Nested Ranker},
	author = {Matveeva, Irina and Burges, Chris and Burkard, Timo and Laucius, Andy and Wong, Leon},
	booktitle = {Proceedings of the 29th Annual International ACM SIGIR Conference on Research and Development in Information Retrieval},
	year = {2006},
	pages = {437--444},
	doi = {10.1145/1148170.1148246},
}

@inproceedings{ceur_overview_longeval_2026,
  author       = {Timo Breuer and
                  Matteo Cancellieri and
                  Alaa El{-}Ebshihy and
                  Maik Fr{\"o}be and
                  Petra Galusc{\'{a}}kov{\'{a}} and
                  Lorraine Goeuriot and
                  Gabriel Iturra{-}Bocaz and
                  J{\"u}ri Keller and
                  Petr Knoth and
                  Andreas Konstantin Kruff and
                  Philippe Mulhem and
                  Florina Piroi and
                  David Pride and
                  Philipp Schaer and
                  Didier Schwab},
  title        = {Extended Overview of the CLEF 2026 LongEval Lab on Longitudinal Evaluation of Model Performance},
  booktitle    = {{CLEF} 2026 Working Notes},
  editor       = {S\'{a}nchez Salido, Eva and Barr\'{o}n-Cede\~{n}o, Alberto and Garc\'{i}a Seco de Herrera, Alba and MacAvaney, Sean and Stru{\ss}, Julia Maria},
  publisher    = {CEUR-WS.org},
  series       = {CEUR Workshop Proceedings},
  year         = {2026},
  pages        = {to appear},
}

\end{document}